\documentclass[8.5pt,twoside,twocolumn]{article}
\usepackage[super,sort&compress,comma]{natbib}
\usepackage{mhchem}
\usepackage{times,mathptmx}
\usepackage{sectsty}
\usepackage{balance}

\usepackage{epsfig,graphicx}
\usepackage{lastpage}
\usepackage[format=plain,justification=raggedright,singlelinecheck=false,font=small,labelfont=bf,labelsep=space]{caption}

\begin{document}

\thispagestyle{plain}

\makeatletter
\def\subsubsection{\@startsection{subsubsection}{3}{10pt}{-1.25ex plus -1ex minus -.1ex}{0ex plus 0ex}{\normalsize\bf}}
\def\paragraph{\@startsection{paragraph}{4}{10pt}{-1.25ex plus -1ex minus -.1ex}{0ex plus 0ex}{\normalsize\textit}}
\renewcommand\@biblabel[1]{#1}
\renewcommand\@makefntext[1]%
{\noindent\makebox[0pt][r]{\@thefnmark\,}#1}
\makeatother
\renewcommand{\figurename}{\small{Fig.}~}
\sectionfont{\large}
\subsectionfont{\normalsize}


\twocolumn[
  \begin{@twocolumnfalse}
\noindent\LARGE{\textbf{Importance of inversion disorder in visible light induced persistent luminescence in Cr$^{3+}$ doped AB$_2$O$_4$ (A = Zn or Mg and B = Ga or Al)}}
\vspace{0.6cm}

\noindent\large{\textbf{Neelima Basavaraju,\textit{$^{a}$} Kaustubh R. Priolkar,$^{\ast}$\textit{$^{a}$} Didier Gourier,\textit{$^{b}$} Suchinder K. Sharma,\textit{$^{b}$} Aur\'{e}lie Bessi\`{e}re,\textit{$^{b}$} and Bruno Viana,\textit{$^{b}$}}}\vspace{0.5cm}

\noindent\textit{\small{\textbf{Received Xth XXXXXXXXXX 20XX, Accepted Xth XXXXXXXXX 20XX\newline
First published on the web Xth XXXXXXXXXX 200X}}}

\noindent \textbf{\small{DOI: 10.1039/b000000x}}
\vspace{0.6cm}

\noindent \normalsize{Cr$^{3+}$ doped spinel compounds AB$_2$O$_4$ with A=Zn, Mg and B=Ga, Al exhibit a long near infrared persistent luminescence when excited with UV or X-rays. In addition, persistent luminescence of ZnGa$_2$O$_4$ and to a lesser extent MgGa$_2$O$_4$, can also be induced by visible light excitation via $^4$A$_2$ $ \rightarrow $ $^4$T$_2$ transition of Cr$^{3+}$, which makes these compounds suitable as biomarkers for in vivo optical imaging of small animals. We correlate this peculiar optical property with the presence of antisite defects, which are present in ZnGa$_2$O$_4$ and MgGa$_2$O$_4$. By using X-ray absorption fine structure (XAFS) spectroscopy, associated with electron paramagnetic resonance (EPR) and optical emission spectroscopy, it is shown that an increase in antisite defects concentration results in a decrease in the Cr-O bond length and the octahedral crystal field energy. A part of the defects are in the close environment of Cr$^{3+}$ ions, as shown by the increasing strain broadening of EPR and XAFS peaks observed upon increasing antisite disorder. It appears that ZnAl$_2$O$_4$, which exhibits the largest crystal field splitting of Cr$^{3+}$ and the smallest antisite disorder, does not show considerable persistent luminescence upon visible light excitation as compared to ZnGa$_2$O$_4$ and MgGa$_2$O$_4$. These results highlight the importance of Cr$^{3+}$ ions with neighboring antisite defects in the mechanism of persistent luminescence exhibited by Cr$^{3+}$ doped AB$_2$O$_4$ spinel compounds.}
\vspace{0.5cm}
 \end{@twocolumnfalse}
  ]

\section{Introduction}
\footnotetext{\dag~Electronic Supplementary Information (ESI) available: [details of any supplementary information available should be included here]. See DOI: 10.1039/b000000x/}


\footnotetext{\textit{$^{a}$Department of Physics, Goa University, Taleigao plateau, Goa 403206, India. Tel: 0832 651 9084; E-mail: krp@unigoa.ac.in}}
\footnotetext{\textit{$^{b}$PSL Research University, Chimie ParisTech - CNRS, Institut de Recherche de Chimie Paris, 75005, Paris, France.}}

Absorption of incident radiation (visible, UV or higher energy) by materials, and delayed subsequent emission most often in the visible range, is termed as long-lasting phosphorescence (LLP) or persistent luminescence. This phenomenon is caused by the trapping of charges (electrons and/or holes) by defects present in the material, preventing fast recombination of the charges. Detrapping of these charges most often proceeds via thermal activation giving rise to a progressive radiative recombination prolonging the emission up to several hours.\cite{1} These materials known as persistent phosphors, were demonstrated to be used as probes for \textit{in vivo} small animal optical imaging in 2007, when emitting in the red or near infrared (NIR) range.\cite{2} This technique is advantageous over conventional fluorescence techniques as it avoids autofluorescence of body tissues under continuous illumination and thus improves the signal to noise ratio.

The technique of \textit{in vivo} imaging was first demonstrated using silicate nanoparticles with composition Ca$_{0.2}$Zn$_{0.9}$Mg$_{0.9}$Si$_2$O$_6$:Eu$^{2+}$, Dy$^{3+}$, Mn$^{2+}$ (CZMSO), with 2.5 mol\% of Mn$^{2+}$ luminescent ion doping.\cite{2} A new compound, Cr$^{3+}$ doped ZnGa$_2$O$_4$ was reported by Bessi\`{e}re et. al. in 2011 to be a potential candidate for this imaging application, with enhanced LLP properties.\cite{3} It was demonstrated in there that LLP results from Cr$^{3+}$ distorted by a neighboring antisite defect, which is the defect resulting from a zinc ion exchanging site position with Ga ion or vice versa.\cite{3} Soon after that, 0.25 mol\% Cr$^{3+}$ doped ZnGa$_2$O$_4$ compound, prepared by hydrothermal method was tested for \textit{in vivo} imaging and was shown to be a suitable biomarker.\cite{4,5} However for solid state synthesis, 0.5 mol\% Cr$^{3+}$ doping was found to be the optimum concentration to get highest LLP.\cite{6} Recently, many other modified gallate spinels doped with Cr$^{3+}$ ions have been discovered to show similar red/NIR LLP emission. Pan et. al. reported Cr$^{3+}$ doped zinc gallogermanates which showed a very long NIR afterglow.\cite{7} An extensive study was done by Allix et. al. on Ge and Sn substituted zinc gallates and they concluded that, substitution of Ga$^{3+}$ by Ge$^{4+}$ or Sn$^{4+}$ in octahedral sites probably increases inversion in the spinel structure, although the substitution may play an extra role in creating more local defects and increase the delayed emission.\cite{8} Lately, Cr$^{3+}$ doped magnesium gallate was also shown to be a good enough phosphor for in vivo imaging with an argument that structural inversion is an important factor governing the LLP property.\cite{9} However, ZnGa$_2$O$_4$ host is a much preferred material to study the LLP mechanism due to its comparatively simpler structure and well resolved Cr$^{3+}$ energy levels.

ZnGa$_2$O$_4$ is known to crystallize in normal spinel structure (cubic space group \textit{Fd3m}) with Zn$^{2+}$ ions in tetrahedral coordination and Ga$^{3+}$ ions in octahedral coordination. However, the existence of a few percent of inversion in site occupancies of Zn and Ga is reported in literature.\cite{3,8,10,11,12} Among the spinel family AB$_2$O$_4$ (with A=Zn, Mg and B=Ga, Al), MgGa$_2$O$_4$ is by and large an inverse spinel compound with about 44\% octahedral sites occupied by Mg$^{2+}$ ions.\cite{9,13} On the other hand, ZnAl$_2$O$_4$ is reported to be a near normal spinel structure with less than 1\% cation disorder.\cite{14} When these spinels are doped with Cr$^{3+}$ in octahedral positions and excited by UV/X-rays, they emit in red/NIR region (around 700 nm). This emission arises due to $^2$E($^2$G) $ \rightarrow $ $^4$A$_2$($^4$F) Cr$^{3+}$ d-d transition.\cite{3,9,15}

Very recently, the LLP mechanism was investigated in detail by thermally stimulated luminescence (TSL) studies done on chromium doped zinc gallate demonstrating that, direct d-d excitation of Cr$^{3+}$ ions by visible light gives a persistent luminescence dominated by Cr$^{3+}$ ions possessing an antisite defect in its first cationic neighbour (referred to as Cr$_{N2}$ ion).\cite{16} The essential feature of the proposed mechanism is that, charge separation and trapping occur in assistance with the local electric field created by the presence of a pair of complementary antisite defects around Cr$^{3+}$ ion. Thus the energy is stored in the form of an electron-hole pair and does not involve valence state change of Cr$^{3+}$ ion. The model efficiently explains how LLP can be excited even with lower energy visible radiation.\cite{16} This hypothesis was further augmented by electron paramagnetic resonance (EPR) studies carried out on chromium doped zinc gallate compounds with varying Zn/(Ga+Cr) nominal ratio.\cite{17} The defects around chromium were identified by correlating photoluminescence (PL) and EPR spectroscopy. Here again it was shown that, Cr$_{N2}$ ion plays a key role in both LLP excitation and emission.\cite{17}

To further substantiate the mechanism deduced from optical and EPR studies, X-ray absorption fine structure spectroscopy (XAFS) measurements were carried out in Cr$^{3+}$ doped ZnGa$_2$O$_4$ to understand the local structures around all the cations. Moreover, the validity of the mechanism is extended by studying local environment of Cr$^{3+}$ in MgGa$_2$O$_4$:Cr$^{3+}$ and ZnAl$_2$O$_4$:Cr$^{3+}$.  Based on interference of  a photoelectron wave emitted from an atom due to absorption of a X-ray photon and a back scattered wave resulting due to its scattering from neighboring ions, XAFS provides information about immediate surroundings of the central absorbing ion. Therefore, XAFS can be used to probe structural defects around a metal ion and is an ideal technique to understand the local structure of an atom in the lattice. It works virtually for all elements and even in cases where the concentration of absorbing ion is very low (few ppm). This makes it a versatile tool to study persistent luminescent materials since the dopant concentration will usually be very low.\cite{18,19,20,21,22,23,24,25,26,27,28} In this paper, we have carried out XAFS measurements at Cr K edges in chromium doped ZnGa$_2$O$_4$, MgGa$_2$O$_4$ and ZnAl$_2$O$_4$ compounds, characterized by optical and EPR spectroscopy. Further, local structures around Zn, Ga and Cr have been studied in Cr$^{3+}$ doped ZnGa$_2$O$_4$ compounds with varying Zn/Ga nominal ratio.

\section{Experimental}

The samples were synthesized by solid state method with their respective metal oxides ZnO (Sigma Aldrich 99.99\% pure), Ga$_2$O$_3$ (Sigma Aldrich 99.99\% pure), MgO (Sigma Aldrich 99.995\% pure), Al$_2$O$_3$ (SRL 99.75\% pure) and CrO$_3$ (SRL 99\% pure) as precursors. Weighed powders along with propan-2-ol were thoroughly mixed in an agate mortar and the dried mixture was pelletized under 4 tons pressure in a hydraulic press. All the pellets were annealed in air at 1300$^\circ$C for 6 hours except ZnAl$_2$O$_4$ which was annealed at 1400$^\circ$C, and later crushed to get fine powders for further characterizations. ZnGa$_2$O$_4$ (noted ZGO) compounds were prepared in three different molar ratios - Zn/Ga = 0.5 (noted s-ZGO for stoichiometric ZGO), Zn/Ga = 0.495 (d-ZGO for 1 mol\% Zn deficiency) and Zn/Ga = 0.505 (e-ZGO for 1 mol\% Zn excess) with 0.5 mol\% Cr introduced relative to Ga. It has to be noted that varying the nominal Zn/Ga ratio in the reactants mix does not mean that Zn or Ga atoms are in excess/deficiency in the obtained ZGO compounds. It was indeed shown that tiny quantities of ZnO/Ga$_2$O$_3$ could either be present as very minor impurities or evaporate during high temperature synthesis.\cite{17} However, varying the nominal Zn/Ga ratio had a very definite effect on the number of point defects in d-ZGO, s-ZGO and e-ZGO compounds.\cite{3,17} Therefore letters d, s and e only reflect deficiency, stoichiometry and excess in the reactants mix without pre-supposing the stoichiometry of the actually formed ZGO compounds. To compare the effect of Cr$^{3+}$ doping on LLP properties in different host lattices, MgGa$_2$O$_4$ (noted MGO) with 1 mol\% Mg deficiency (noted d-MGO) and 0.5 mol\% Cr doping relative to Ga, and ZnAl$_2$O$_4$ (noted ZAO) with 1 mol\% Zn deficiency (noted d-ZAO) and 0.5 mol\% Cr doping relative to Al, were synthesized.

Room temperature (RT) photoluminescence (PL) excitation spectra were recorded on Varian Cary Eclipse spectrofluorimeter in the range 190 nm-650 nm with xenon lamp as excitation source. Pulsed laser excited PL was run on 8 mm-diameter pellets silver glued on the cold finger of a cryogenic system maintained at 20 K. The emitted light was collected by an optical fiber and transmitted to a Scientific Pixis 100i CCD camera cooled at -20$^\circ$C and coupled to a monochromator with 1200 groves/mm grating. The pellets were excited at 230 nm by an optical parametric oscillator (OPO) EKSPLA NT342B. The PL emission spectra were measured with 10 ms gate width and 26 ns gate delay. LLP measurements were carried out at RT on 180 mg samples filled into a 1 cm diameter circular sample holder. The samples were illuminated for 15 minutes with X-rays (Mo-tube, 20 mA-50 kV) or for 30 minutes in the $^4$A$_2$ $ \rightarrow $ $^4$T$_2$ Cr$^{3+}$ band with OPO. After this excitation, emission was collected using a Scientific Pixis 100 CCD camera via an optical fiber linked to an Acton SpectraPro 2150i spectrometer for spectral analysis. LLP emission spectra were recorded during the excitation and 5 s after the end of excitation. All the samples were bleached at 250$^\circ$C for 30 minutes and were kept in the dark prior to LLP measurements.

X-band ($\sim$9.4 GHz frequency) EPR spectra were recorded at room temperature on weight normalized samples using a Bruker Elexsys E500 continuous wave spectrometer. The powder spectra were simulated with the X-Sophe software tool from Bruker.

A Rigaku X-ray diffractometer was used to obtain the X-ray diffraction (XRD) patterns at RT using Cu-K$\alpha$ radiation. The spectra were recorded in 2$\theta$ range 20$^\circ$-80$^\circ$ with 0.02$^\circ$ step and 2$^\circ$/min scan speed. Rietveld refinement on the XRD patterns was carried out using FullProf software.\cite{29} XAFS at RT were measured on the samples in fluorescence mode for Cr K edge and in transmission mode for Zn K and Ga K edges, at SAMBA beamline in Soleil synchrotron facility, France. Si (111) crystal plane was used as the monochromator. For fluorescence measurements, absorbers were prepared by mixing 50 mg compound with 100 mg boron nitride and pressing each of them into 10 mm pellets while for transmission, the appropriate amount of finely ground powder was deposited on a membrane. Fluorescence yield was collected via Canberra 35 pixels SSD detector and transmitted photons were counted using ionization chamber with appropriate gases. Extended X-ray absorption fine structure spectroscopy (EXAFS) fitting was carried out using Ifeffit software with Athena and Artemis programs.\cite{30}  EXAFS data in the k range of 2 to 11 \AA$^{-1}$ for Zn, 2 to 14 \AA$^{-1}$ for Ga and 3 to 10 \AA$^{-1}$ for Cr K edge were Fourier transformed, and the fitting was performed in the R range of 1 to 3.6 \AA ~to obtain reasonable fits. Theoretical amplitude and phase information for various scattering paths were obtained using FEFF 6.01\cite{31} and the Rietveld refined parameters.

\section{Results}

\begin{figure}
\centering
\includegraphics[width=\columnwidth]{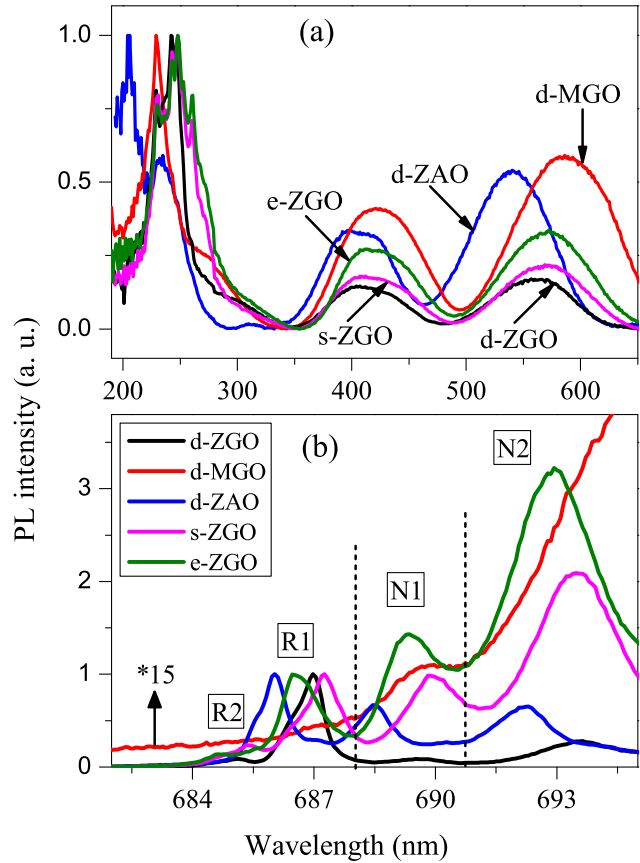}
\caption{(a) PL excitation spectra of Cr$^{3+}$ doped ZGO compounds with varying Zn/Ga nominal ratio, d-MGO and d-ZAO compounds, measured at room temperature. (b) Corresponding Zero-phonon PL emission spectra at 20 K and excited at 230 nm.}
\label{fig:pl}
\end{figure}

Photoluminescence excitation spectra measured at RT for s-ZGO, e-ZGO, d-ZGO, d-MGO and d-ZAO are presented in Figure \ref{fig:pl} (a). The spectra consist of host band gap excitation peaking at about 245 nm for ZGO compounds, 230 nm for d-MGO and 205 nm for d-ZAO, and three broad absorption bands around 230-270 nm, 410 nm and 560 nm belonging to $^4$A$_2$($^4$F) $ \rightarrow $ $^4$T$_1$($^4$P), $^4$A$_2$($^4$F) $ \rightarrow $ $^4$T$_1$($^4$F) and $^4$A$_2$($^4$F) $ \rightarrow $ $^4$T$_2$($^4$F) Cr$^{3+}$ d-d transitions respectively. It was previously shown that, crystal field around Cr$^{3+}$ ion in d-MGO is weaker than that of ZGO compounds and hence a red shift is seen in Cr$^{3+}$ absorption bands of d-MGO compared to ZGO.\cite{9} On the contrary in d-ZAO, Cr$^{3+}$ absorption bands are shifted to shorter wavelengths compared to ZGO compounds indicating that the crystal field around Cr$^{3+}$ ion is stronger than that of ZGO. Within the ZGO compounds, the Cr$^{3+}$ absorption bands are seen to be gradually moving towards longer wavelengths with the increasing Zn/Ga nominal ratio (deficiency to stoichiometric to excess) revealing a decrease in the crystal field around Cr$^{3+}$ ion.

The influence of defects around Cr$^{3+}$ ions among different samples could also be seen in PL emission at 20 K under 230 nm excitation (Figure \ref{fig:pl} (b)). The emission lines correspond to typical $^2$E $ \rightarrow $ $^4$A$_2$ transitions of Cr$^{3+}$ with either unperturbed environment and/or perturbation from the nearby crystal defects. Only zero phonon lines (ZPL) are shown in Figure \ref{fig:pl} (b). Multi phonon side bands (PSB) which occur at shorter wavelengths, are not shown here. For the sake of comparison, the spectra except for d-MGO sample are normalized to R1 line. The d-MGO spectrum is amplified by a factor of 15 with respect to d-ZGO, for better visibility and comparison. The emission spectra can be categorized into three main ZPL regions, namely, (i) R1 and R2 region; (ii) N1 region and (iii) N2 region, separated by dotted lines in Figure \ref{fig:pl} (b). R1 and R2 lines correspond to the Cr$^{3+}$ ions with an unperturbed ideal environment (referred to as Cr$_R$ centers). The distinct R1 and R2 lines are ascribed to the splitting of $^2$E excited state of Cr$^{3+}$ due to trigonal distortion into two levels separated by $\sim$40 cm$^{-1}$ in ZGO and $\sim$7 cm$^{-1}$ in ZAO.\cite{32,33} Similarly, the N1 line (referred to as Cr$_{N1}$) may correspond to either a Cr$^{3+}$-$V_{Zn}$ pair ($V_{Zn}$ is a Zn vacancy)\cite{15} or a Cr$^{3+}$-$Zn_i$ pair ($Zn_i$ is an interstitial Zn)\cite{34} or a Cr$^{3+}$ close to an antisite defect\cite{17,33}. The N2 line is assigned to the presence of antisite defects (presumably ${Zn_{Ga}}'$) close to Cr$^{3+}$ ion as first cationic neighbour (referred to as Cr$_{N2}$).\cite{17,33,35,36} The emission wavelengths of all these lines are reported in Table \ref{tbl:pl}. As N lines are due to Cr$^{3+}$ perturbed by neighboring antisite defects, their intensity (compared to R lines) increases with increasing inversion disorder. This shows that d-ZAO and d-ZGO are the less disordered materials, followed by s-ZGO and e-ZGO in increasing order. The disorder is so high in d-MGO that emission features are difficult to interpret and the main emission line is situated at 707 nm with total of six ZPLs revealing up to six different environments for Cr$^{3+}$ ions.\cite{9}

\begin{table}[h]
\caption{Characteristics of zero-phonon emission lines (ZPL) of Cr$^{3+}$-doped ZGO, ZAO and MGO compounds.}
\label{tbl:pl}
\begin{center}
\begin{tabular}{l  c c c c}
\hline\hline
Sample & \multicolumn{4}{c}{ZPL emission line (682-695nm)} \\
       &  R1 (nm)   &  R2 (nm)   &  N1 (nm)   &  N2 (nm)   \\ [0.5ex]
 \hline
d-ZAO  &  686.0   &   684.2  &  688.5   &  692.2   \\ [0.5ex]
d-ZGO  &   686.4  &  684.6   &   689.3  &  692.9   \\ [0.5ex]
s-ZGO  &   686.9  &  685.1   &   689.7  &  693.4   \\ [0.5ex]
e-ZGO  &   687.2  &  685.4   &   689.9  &  693.6   \\ [0.5ex]
d-MGO  &   N. V.  &  N. V.   &   N. V.  &  N. V.   \\ [0.5ex]
                \hline\hline
\end{tabular}
\end{center}
N. V. - Not visible
\end{table}

\begin{figure}
\centering
\includegraphics[width=\columnwidth]{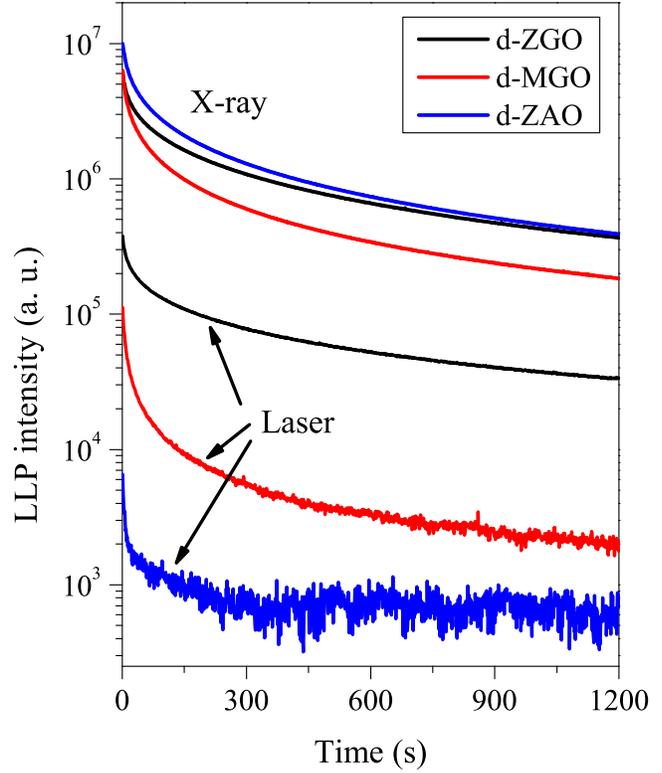}
\caption{LLP decay curves of Cr$^{3+}$-doped compounds after 15 minute X-ray irradiation and after 30 minute laser excitation of wavelength 560 nm for d-ZGO, 585 nm for d-MGO and 540 nm for d-ZAO samples. These wavelengths correspond to the maximum of the $^4$A$_2$($^4$F) $ \rightarrow $ $^4$T$_2$($^4$F) absorption band of Cr$^{3+}$.}
\label{fig:llp}
\end{figure}

The persistent luminescence decay curves obtained with X-ray and laser (wavelength at the maximum of $^4$A$_2$($^4$F) $ \rightarrow $ $^4$T$_2$($^4$F) Cr$^{3+}$ d-d transition) excitations are presented in Figure \ref{fig:llp}. d-MGO is known to show less LLP intensity compared to ZGO with both X-ray and laser excitations as reported earlier.\cite{9} This is presumably due to excess cationic inversion which partially quenches LLP by introducing a direct recombination pathway between abundant defects.\cite{9} The d-ZAO compound which is a near normal spinel, shows almost same or slightly better LLP intensity as compared to d-ZGO compound with X-ray excitation. However, d-ZAO shows very feeble LLP intensity with 540 nm laser excitation, although this excitation corresponds to the maximum of the $^4$A$_2$($^4$F) $ \rightarrow $ $^4$T$_2$($^4$F) absorption band of Cr$^{3+}$. On the other hand, d-ZGO shows consistently better LLP with both excitations as seen in Figure \ref{fig:llp}. Also, this distinctive property of d-ZGO to show LLP with laser excitation renders the possibility for re-excitation of the ZGO nanoparticles inside the animal body by visible light illumination, making it a favourable candidate for the application of in vivo imaging.\cite{4,5}

\begin{figure}
\centering
\includegraphics[width=\columnwidth]{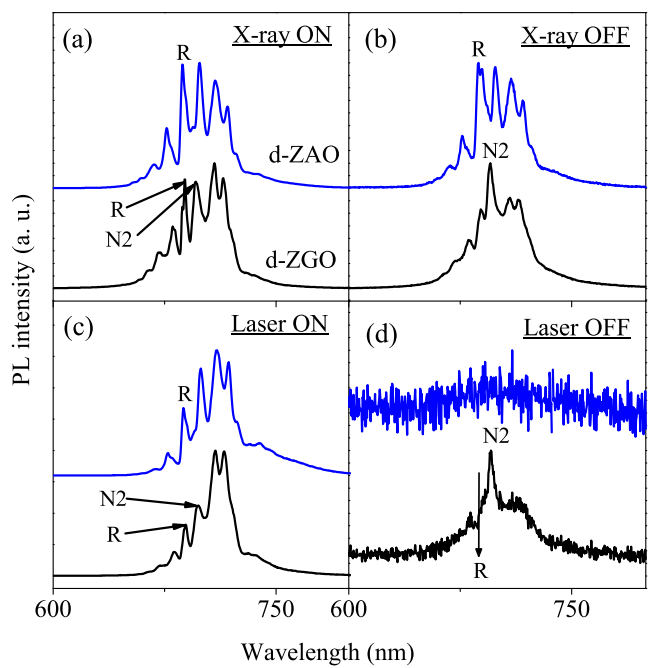}
\caption{Emission spectra measured for d-ZGO (black curves) and d-ZAO (blue curves) (a) during X-ray irradiation; (b) 5 s after the end of 15 minute X-ray illumination; (c) during optical laser excitation at 540 nm for d-ZAO and 560 nm for d-ZGO; (d) 5 s after the end of 30 minute laser excitation. Position of R line is indicated by an arrow.}
\label{fig:llpem}
\end{figure}

Comparison of emission spectra recorded for d-ZGO and d-ZAO compounds during the excitation and during LLP emission (after the end of excitation) are shown in Figure \ref{fig:llpem}. Emission spectra measured during X-ray excitation (Figure \ref{fig:llpem} (a)) show intense R-line for both the compounds, and corresponds to the $^2$E $ \rightarrow $ $^4$A$_2$ ZPL of Cr$^{3+}$. This ZPL is flanked by multi phonon sidebands. N2 line which arises due to emission from Cr$^{3+}$ ions with an antisite defect in its first cationic neighbour is clearly seen in d-ZGO spectrum. It is difficult to identify the presence of N1 and N2 in d-ZAO because of the intense multi PSB in the same range. However low temperature emission spectra (Figure \ref{fig:pl} (b)) without PSB clearly shows the presence of N1 and N2 lines. The X-ray excited LLP emission spectrum for d-ZAO shows an intense R-line (which is weakly split) whereas d-ZGO exhibits a prominent N2 line (Figure \ref{fig:llpem} (b)). This indicates that for X-ray excitation, the charge recombination is taking place mainly at undistorted Cr$^{3+}$ ions (Cr$_R$) in d-ZAO, while it occurs through Cr$_{N2}$ in d-ZGO. With laser excitation, the emission spectra recorded while the laser is on (Figure \ref{fig:llpem} (c)) show R and N2 lines for both compounds. When the laser is switched off (LLP emission), d-ZAO shows a broad weak emission (Figure \ref{fig:llpem} (d)), whereas d-ZGO spectrum again exhibits a prominent N2 line. This indicates the crucial role of Cr$_{N2}$ ions and antisite defects in LLP emission when excited in low energy absorption bands of Cr$^{3+}$.

\begin{figure}
\centering
\includegraphics[width=\columnwidth]{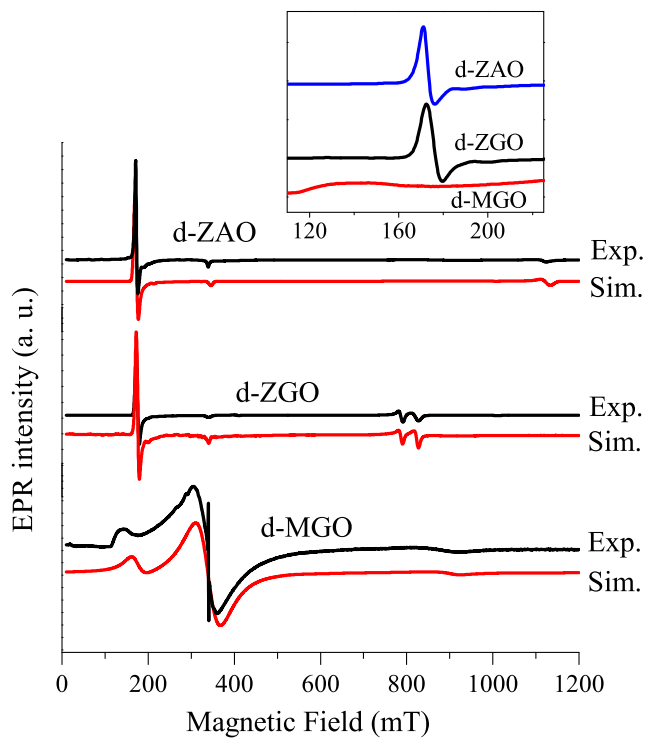}
\caption{X-band EPR spectra of d-ZAO, d-ZGO and d-MGO recorded at room temperature. Microwave power 2 mW; Modulation depth 1 mT at 100 kHz modulation frequency. The insert shows a zoom of the low magnetic field part of the spectra.}
\label{fig:epr}
\end{figure}

Experimental and simulated EPR spectra of Cr$^{3+}$ in d-ZAO, d-ZGO and d-MGO are reported in Figure \ref{fig:epr}. For the direct spinels d-ZGO and d-ZAO, which exhibit a weak inversion disorder, the spectrum is composed of a strong line around 175 mT, a weak line around 330-350 mT and other lines at $\sim$800 mT in d-ZGO and $\sim$1100 mT in d-ZAO. These spectra were simulated with EPR parameters of a Cr$^{3+}$ ion in a weakly axially distorted octahedral site, in agreement with the C3 symmetry of Ga and Al sites in these compounds. The effect of antisite defects is a strain broadening of these EPR lines, which increases with magnetic field strength so that EPR lines at high magnetic field are no longer observed in s-ZGO and e-ZGO.\cite{17} A zoom of transition at the low magnetic field (175 mT) shows that the line is slightly broader in d-ZGO than in d-ZAO, pointing to a slightly higher disorder in d-ZGO than in d-ZAO. The inverse spinel d-MGO gives a very different EPR spectrum. We still recognize broad lines at $\sim$130-150 mT (see also the insert in Figure \ref{fig:epr}) and at $\sim$900 mT attributable to Cr$^{3+}$ in octahedral sites affected by important disorder due to abundant antisite defects. The spectrum also exhibits a broad and symmetrical line in the field range $\sim$300 mT. This line was also observed in disordered s-ZGO and e-ZGO materials.\cite{17} Based on the temperature dependence of its intensity, this line was tentatively attributed to clusters of antiferromagnetically coupled Cr$^{3+}$ ions.\cite{17} The decomposition of the simulated spectrum of Cr$^{3+}$ in d-MGO is shown in Supplementary Figure 1, and the simulation parameters of all compounds are reported in Supplementary Table 1. The parameter which controls the shape of the powder EPR spectrum of Cr$^{3+}$ in these compounds is the zero field splitting (ZFS) parameter \lq D\rq. It represents the splitting 2D of the $^4$A$_2$ ground state by the combined effect of the spin-orbit coupling and the trigonal distortion of the octahedral crystal field. d-ZAO exhibits the largest splitting (2D = 1.864 cm$^{-1}$) compared to d-MGO (2D = 1.270 cm$^{-1}$) and d-ZGO (2D = 1.050 cm$^{-1}$). This corresponds to a decreasing trigonal distortion in the sequence d-ZAO$\>>\>>$d-MGO$\>>$d-ZGO, which can be compared to octahedral crystal field splitting decreasing in the sequence d-ZAO$\>>$d-ZGO$\>>$d-MGO (see Figure \ref{fig:pl} (a)). Thus the aluminate d-ZAO has more crystal field splitting D, more trigonal distortion and less antisite disorder than gallates d-ZGO and d-MGO.

To complement optical and EPR studies, and to investigate in more details the environment around Cr$^{3+}$ ion in these lattices, structural studies including XRD and EXAFS measurements have been carried out. XRD patterns of all the compounds s,d,e-ZGO, d-MGO and d-ZAO are shown in Supplementary Figure 2 and Rietveld refinement parameters are given in Supplementary Table 2. XRD indicates the formation of pure cubic spinel compounds. As previously reported, a minor $\sim$1\% MgO impurity in d-MGO compound (marked + in Supplementary Figure 2) is identified.\cite{9} A clear shift in the peak positions is observed in Cr$^{3+}$ doped host lattices spectra, corresponding to the variation in lattice parameters which is mostly the result of difference in ionic sizes of cations. The lattice parameter decreases in the sequence ZGO$\>>$d-MGO$\>>$d-ZAO. Not much variation was visibly seen among ZGO compounds however, Rietveld refinement indicated lowering lattice constant values with increasing Zn/Ga nominal ratio. Rietveld refinement was done on all the XRD patterns with A site cations in tetrahedral 8a positions (site symmetry: T$_d$) and B site cations in octahedral 16d positions (site symmetry: D$_{3d}$).\cite{37} Parameters like lattice constant, cationic site occupancy along with scale factor, background and instrumental parameters were varied in fitting. The fits to the experimental patterns along with residues are shown in Supplementary Figure 2. The cationic occupancy could not be varied for ZGO compounds since Zn$^{2+}$ and Ga$^{3+}$ ions are isoelectronic having similar X-ray scattering powers for the Cu-K$\alpha$ radiations. As reported earlier, d-MGO compound shows 45.2(3)\% cationic site inversion confirming the near inverse spinel structure.\cite{9} Site occupancy was also varied for d-ZAO compound and yielded no inversion in the lattice hinting towards its normal spinel crystal structure.

\begin{figure}
\centering
\includegraphics[width=\columnwidth]{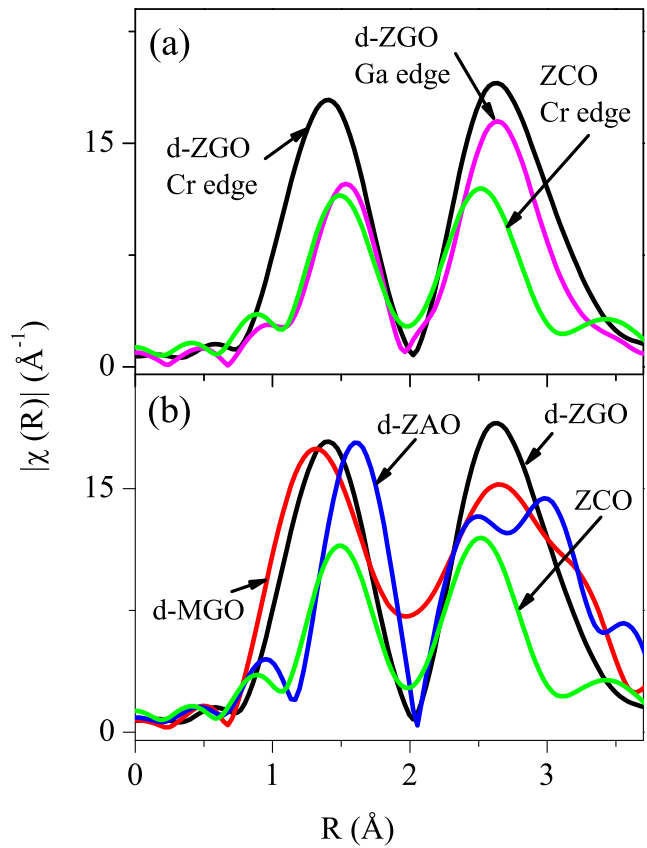}
\caption{Fourier transform magnitude of EXAFS pattern in the k range 3 to 10 \AA$^{-1}$ (a) of Cr K edge and Ga K edge for d-ZGO and ZCO; (b) of Cr K edge for d-ZGO, d-MGO, d-ZAO and ZCO.}
\label{fig:exafs1}
\end{figure}

EXAFS measurements were carried out on Cr$^{3+}$ doped ZGO, d-MGO and d-ZAO compounds, along with ZnCr$_2$O$_4$ (noted ZCO) compound which was used as a reference for octahedral Cr$^{3+}$ environment. Magnitude of Fourier transform (FT) of Cr K and Ga K edge EXAFS in d-ZGO are presented in Figure \ref{fig:exafs1} (a) in comparison with Cr K edge spectrum of ZCO. The Cr K EXAFS spectra in all three ZGO compounds are largely similar and hence the individual spectra are not presented here. It can be seen that the first peak corresponding to Cr-O correlation in d-ZGO Cr edge appears at a lower R value compared to the first peak in both d-ZGO Ga edge and ZCO Cr edge, which implies that the average Cr-O distance in d-ZGO is less than the Ga-O distance in the same compound or the Cr-O distance in ZCO. This anomaly gains more importance because unlike the first peak, the second intense peak observed around 2.7 \AA ~in d-ZGO Cr edge spectrum is almost at the same position as that in d-ZGO Ga edge spectrum. Also, an asymmetrical peak broadening of the first peak is observed in d-ZGO Cr edge spectrum revealing a distribution of Cr-O bond distances, especially towards the shorter distances. These observations indicate that the local octahedral environment around Cr$^{3+}$ ion is largely distorted compared to that of Ga$^{3+}$ environment in ZGO or the ideal octahedral Cr$^{3+}$ environment in ZCO. FT magnitude of k weighted EXAFS spectra of Cr$^{3+}$ doped host lattices are presented in Figure \ref{fig:exafs1} (b). A comparison of the FT magnitudes in the host lattice compounds with that of ZCO shows Cr-O distances in d-ZGO and d-MGO to be shorter than that in ZCO. Again the peaks are asymmetrically broadened indicating a distribution in Cr-O bond distances. However in d-ZAO, Cr-O distance is larger than that in ZCO and no broadening is seen in the peak indicating a smaller distribution of Cr-O distances.

Experimentally obtained EXAFS spectra were fit using Ifeffit software for all the compounds to obtain bond distances (R) and mean-square disorders ($\sigma^2$) for each path. The raw spectra were background corrected and energy calibrated using Athena and then fitted using Artemis. Atomic coordinates and lattice parameters obtained for each compound from Rietveld refinement were used as inputs to generate a FEFF input file, with Cr or Zn or Ga as core for the corresponding Cr or Zn or Ga K edge spectra. The photoelectron scattering paths were then calculated and the experimental data was fitted up to 3.6 \AA ~in \textit{R}-space to obtain R and $\sigma^2$ for each scattering path. Goodness of the fit, expressed by \textit{R factor}, was less than 0.03 in all the fits. The Cr K EXAFS in d-ZGO compound could only be fitted using the third cumulant parameter ($C_3$) which indicates an asymmetrical distribution of Cr-O bond distances. All other attempts to fit the spectra resulted in negative (unphysical) values of $\sigma^2$. This deviation away from a normal distribution of Cr-O bond distances can be explained to be due to the disorder around the Cr$^{3+}$ ion, resulting in an asymmetric distribution of bond distances. An example of fit to EXAFS data in back transformed \textit{k} space is presented in Figure \ref{fig:exafs2} for d-ZGO. Other fits for Zn K edge and Ga K edge EXAFS patterns of d-ZGO are shown in Supplementary Figure 3. All the fitting parameters are given in Supplementary Tables 3 for Cr K edge EXAFS and Table 4 for Zn K and Ga K EXAFS.

\begin{figure}
\centering
\includegraphics[width=\columnwidth]{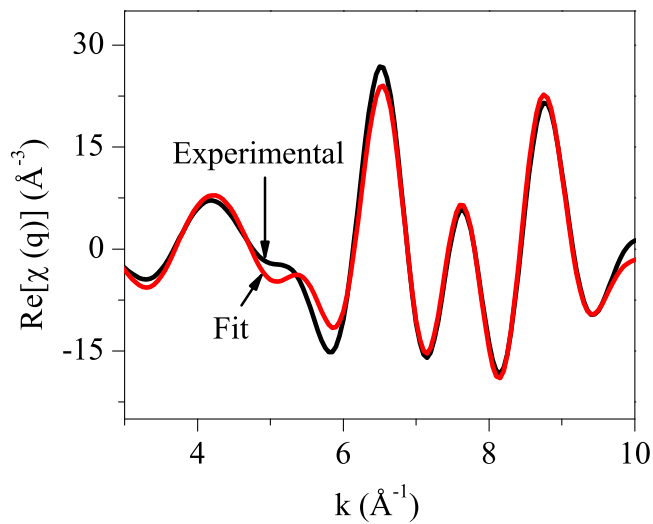}
\caption{Experimentally obtained Cr K edge EXAFS pattern for d-ZGO along with fit in \textit{q} space in the k range 3 to 10 \AA$^{-1}$.}
\label{fig:exafs2}
\end{figure}

Unlike ZGO compounds, Cr K EXAFS in d-MGO could be fitted starting with two structural models based on normal spinel structure with Mg occupying all tetrahedral sites and Ga occupying octahedral sites, and inverse spinel structure wherein all the Mg ions occupied octahedral sites and half of the Ga ions occupied the tetrahedral sites. Phase fraction of each model was taken as a fitting parameter. The inversion around Cr$^{3+}$ for d-MGO was found to be 44(5)\% which is consistent with the value 45.2(3)\% obtained from Rietveld refinement. In case of d-ZAO, best fit was obtained with the normal spinel model, indicating presence of very little or no inversion around Cr$^{3+}$ ion.

\section{Discussion}

\begin{figure}
\centering
\includegraphics[width=\columnwidth]{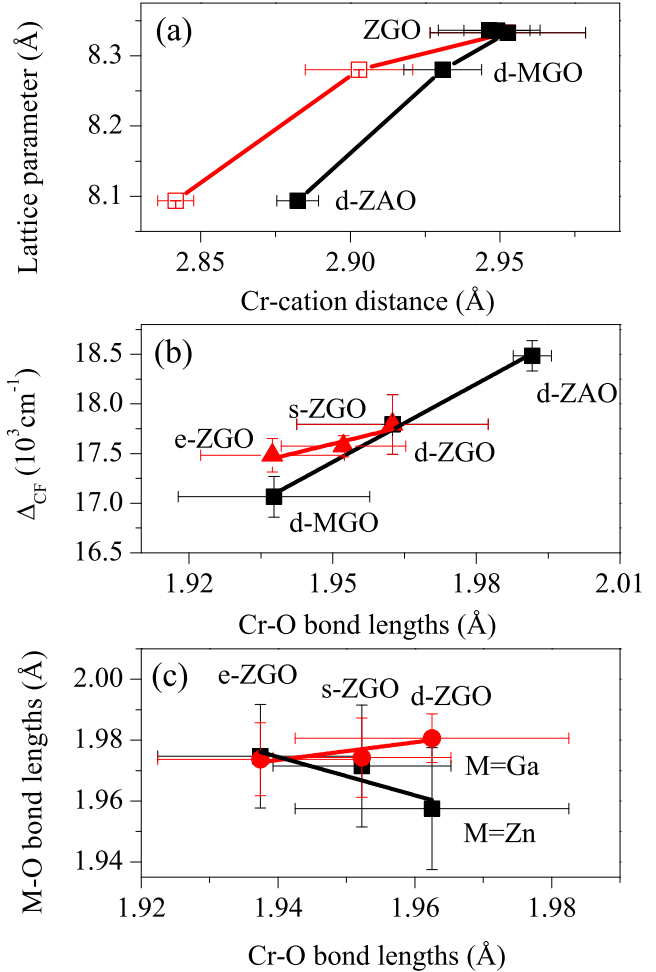}
\caption{(a) Rietveld refined lattice parameter values plotted versus Cr-first cationic neighbour distances obtained by EXAFS fitting. Filled black squares represent Cr-trivalent ion distances and open red squares represent Cr-divalent ion distances. The three ZGO compounds (d-ZGO, s-ZGO and e-ZGO) have almost same values and they are seen to be overlapping. (b) Crystal field energy $\Delta_{cf}$ calculated from $^4$A$_2$ $ \rightarrow $ $^4$T$_2$ Cr$^{3+}$ absorption band in RT PL spectra versus Cr-O bond lengths obtained from EXAFS fitting. (c) Metal-oxygen bond lengths (with M=Ga, Zn) plotted versus Cr-O bond lengths obtained from EXAFS fitting for ZGO compounds. Filled squares represent Zn-O bond lengths and filled circles represent Ga-O bond lengths.}
\label{fig:corre}
\end{figure}

This work contributes to our effort to explain an unexpected property of Cr$^{3+}$-doped ZGO, which is the possibility to activate LLP by visible light. Owing to the leading role of Cr$^{3+}$ ions in this property, we examined the perturbation of Cr$^{3+}$ environment by structural defects, mainly antisite defects due to inversion disorder. By combining PL, TSL and EPR analyses, we recently proposed a mechanism whereby LLP excitation and emission are likely due to a particular type of [Cr$^{3+}$-defect] cluster (Cr$_{N2}$), namely a Cr$^{3+}$ ion with two neighboring antisite defects of opposite charge (a ${Zn_{Ga}}'$ defect at short distance and a ${Ga_{Zn}}^\circ$ defect at longer distance).\cite{16,17} In this mechanism, it was proposed that the electric field created by the two neighboring defects of opposite charge triggers the formation of electron-hole pairs from the excited $^4$T$_2$ chromium state (or other states of higher energy), the hole and the electron being next trapped at ${Zn_{Ga}}'$ site and ${Ga_{Zn}}^\circ$ site of the lattice, respectively. The present structural study provides a deeper insight into this still speculative model. The importance of antisite defects is now clearly supported by the correlation between the amount of inversion disorder and the excitation of LLP by visible light. ZGO is characterized by a few percent of inversion ($\sim$3\%),\cite{8} which appears a good compromise for this optical property. On the contrary d-ZAO exhibits no measurable inversion and no strain broadening of EPR, PL and XAFS spectra, and clearly shows no considerable visible light excited LLP, although it still exhibits a strong X-ray excited LLP (Figure \ref{fig:llp}). Indeed TSL studies on this compound clearly show the lack of TSL peak arising due to antisite defects (occurring at 370 K).\cite{38} d-MGO shows a high level of inversion disorder ($\sim$45\%) which manifests itself by strong strain broadening effects in EPR, PL and XAFS spectra. This compound presents a less efficient visible light excited LLP than d-ZGO, presumably due to a parasitic direct recombination of trapped charges (Figure \ref{fig:llp}). Antisite defects are thus crucial to allow the excitation, charge trapping and emission of the LLP, when excited by visible light. The effect of these cationic defects and more generally, of the nature of cations on Cr$^{3+}$ environment, is shown in Figure \ref{fig:corre} which correlates interatomic distances, lattice parameters and crystal field energies. It appears that Cr-Zn (or Cr-Mg) distances (open squares in Figure \ref{fig:corre} (a)) and Cr-Ga (or Cr-Al) distances (full squares in Figure \ref{fig:corre} (a)) are entirely determined by lattice parameter (see also Supplementary Tables 3 and 4). The presence of defects does not affect this correlation as shown by the fact that d-ZGO, s-ZGO and e-ZGO have almost the same lattice parameters and the same Cr-Ga distances of 2.95 \AA.

\begin{figure}
\centering
\includegraphics[width=\columnwidth]{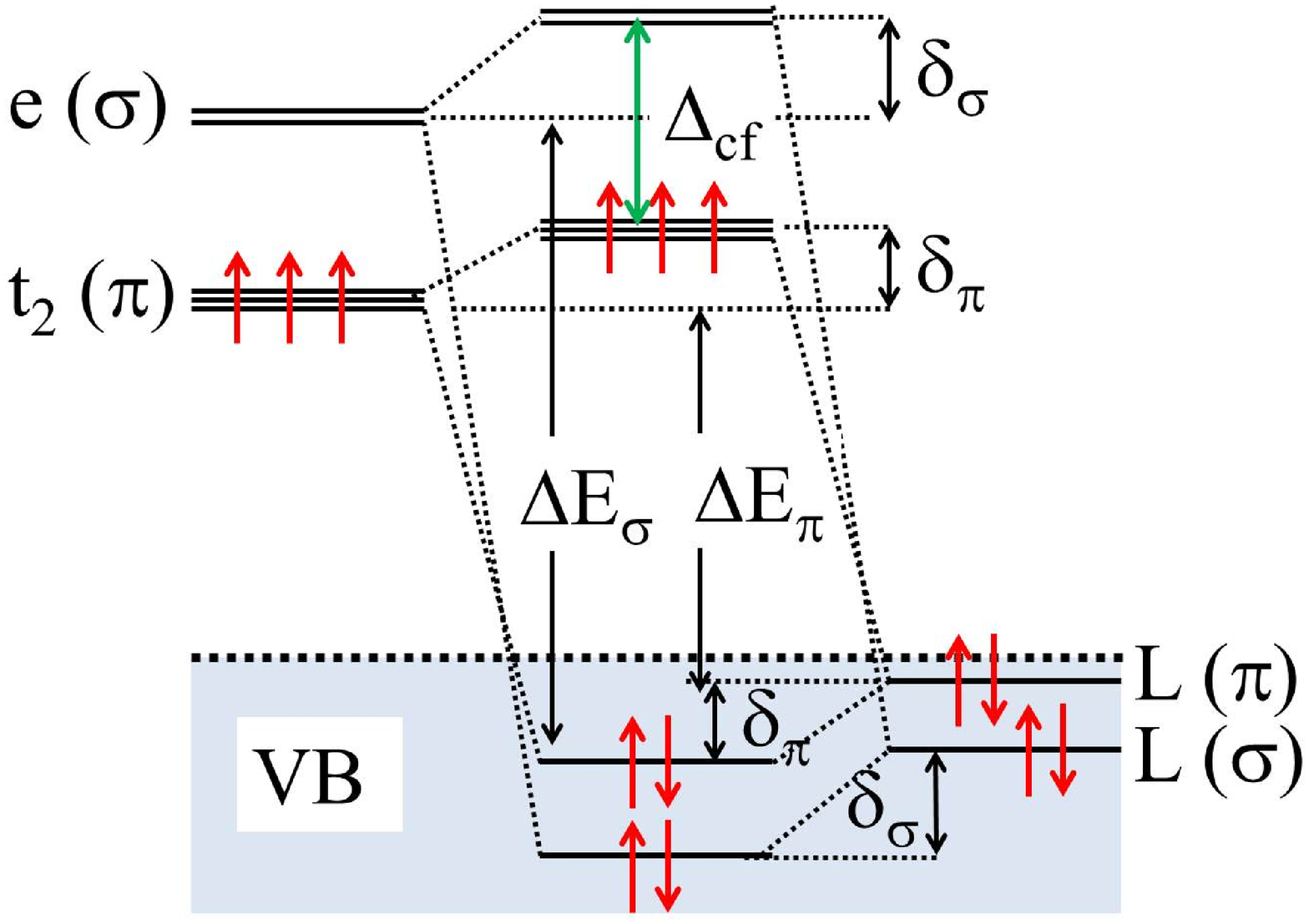}
\caption{Energy level scheme representing the effect of $\sigma$ and $\pi$ contributions to Cr-O bond.}
\label{fig:crystf}
\end{figure}

Contrary to Cr-cation distances, there is no correlation between Cr-O bond lengths and lattice parameters (see Supplementary Table 3). However a linear correlation is clearly observed in Figure \ref{fig:corre} (b) between Cr-O bond length and the crystal field energy $\Delta_{cf}$ deduced from the $^4$A$_2$ $ \rightarrow $ $^4$T$_2$ absorption band of Cr$^{3+}$, which occurs at energy $h\nu$ = $\Delta_{cf}$. This correlation and the lack of correlation with lattice parameter show that Cr$^{3+}$ imposes its first shell environment. This can be understood by considering the relation between the strength of the Cr-O bond and the crystal field splitting energy as shown in Figure \ref{fig:crystf}. In a purely ionic model and ignoring for the moment the electron repulsion, the electronic configuration of Cr$^{3+}$ in octahedral environment is t$_2^3$e$^0$ with a crystal field splitting $\Delta_{cf}$ between t$_2$ and e levels. Oxygen 2p orbitals participate to the top of the valence band (L in Figure \ref{fig:crystf}). The t$_2$ and e metal orbitals are separated from the valence band by energies of the order of $\Delta$E$_\pi$ and $\Delta$E$_\sigma$, respectively. The $\pi$ and $\sigma$ covalent characters of the Cr-O bond are described by non-zero transfer integrals $\beta_\pi$ = $\langle$t$_2\vert$\textit{H}$\vert$L($\pi$)$\rangle$ and $\beta_\sigma$ = $\langle$e$\vert$\textit{H}$\vert$L($\sigma$)$\rangle$ between metal 3d and oxygen 2p orbitals, where \textit{H} is the Hamiltonian operator and L($\pi$) and L($\sigma$) are the symmetry adapted linear combinations of oxygen orbitals with $\pi$ and $\sigma$ character, respectively. These transfer integrals are proportional to metal-ligand overlap. A strengthening of the covalent bonding (decrease of the Cr-O bond length) will manifest itself by a shift $\sim\delta$ to low energy for L orbitals (bonding character) and a shift $\sim\delta$ to high energy for metal orbitals (antibonding character). An increase of the $\pi$ character of the Cr-O bond will thus decrease the crystal field splitting $\Delta_{cf}$ by an amount $\delta_\pi$ $\approx$ 2$\beta^2_\pi$/$\delta$E$_\pi$. Alternatively, an increase of the $\sigma$ character of the bond will increase $\Delta_{cf}$ by an amount $\delta_\sigma$ $\approx$ 2$\beta^2_\sigma$/$\delta$E$_\sigma$ (Figure \ref{fig:crystf}). Within this scheme, the decrease of Cr-O bond length observed in the sequence $d_{Cr-O}$(d-ZAO) $>$ $d_{Cr-O}$(d-ZGO) $>$ $d_{Cr-O}$(d-MGO) (Figure \ref{fig:corre} (b)) clearly corresponds to a decrease of $\Delta_{cf}$, which points to an increase of the $\pi$ contribution to the Cr-O bond along this series. Alternatively, an increase of the $\sigma$ contribution to the Cr-O bond would produce the opposite correlation, i.e. an increase of $\Delta_{cf}$ associated to a decrease of Cr-O bond length.

The same type of correlation, with a smaller slope, was also observed by varying the defect concentration in ZGO (Figure \ref{fig:corre} (b)). As reported before, the concentration of defects around Cr$^{3+}$ is minimum in d-ZGO and increases in the sequence d-ZGO $<$ s-ZGO $<$ e-ZGO.\cite{3,17} The fact that the slope is smaller upon varying the defect concentration (full triangles in Figure \ref{fig:corre} (b)) than upon varying the A$^{2+}$ and B$^{3+}$ cations (full squares in Figure \ref{fig:corre} (b)) can be explained by symmetry effects. In the spinel series AB$_2$O$_4$ (A=Mg, Zn and B=Al, Ga), the symmetry of the unperturbed Cr$^{3+}$ site is trigonal, which introduces no mixing between t$_2$($\pi$) orbitals and e($\sigma$) orbitals. An antisite defect in first neighbour position of Cr$^{3+}$, and located outside the C3 axis, decreases the site symmetry which becomes non axial.\cite{17} This introduces a mixing between t$_2$ and e orbitals, so that t$_2$ orbitals take a small $\sigma$ character while e orbitals take a small $\pi$ character. As $\sigma$ bonds are stronger than $\pi$ bonds, this increasing $\sigma$-$\pi$ mixing upon increasing the defect concentration shortens the Cr-O bond length. However, both t$_2$ and e orbital sets are now shifted to high energy, so that the increase of $\Delta_{cf}$ is smaller than for a pure axial symmetry, explaining the smaller slope in the curve $\Delta_{cf}$ = $f$($d_{Cr-O}$) in the ZGO series. Introducing the electron repulsion between the three t$_2$ electrons does not modify this scheme. The $^2$E state (emitting level) and the $^4$A$_2$ state (ground state level) both correspond to the configuration t$_2^3$ with total spin S=1/2 and S=3/2, respectively. The effect of the crystal field variation is much smaller for the $^2$E-$^4$A$_2$ splitting than for the $^4$T$_2$-$^4$A$_2$ splitting, but varies with the same trend.\cite{39} This effect thus explains why lower the symmetry of the Cr$^{3+}$ site, smaller the Cr-O bond length, and lower the energy of the $^2$E-$^4$A$_2$ emission of Cr$^{3+}$. This confirms the correlation previously proposed between the decreasing symmetry of the [Cr$^{3+}$-defect] clusters determined by EPR (Cr$_R$ $>$ Cr$_{N1}$ $>$ Cr$_{N2}$) and the decreasing photon energy of the $^2$E-$^4$A$_2$ emission line.\cite{17} Clusters with lowest symmetry give N2 emission line at the lowest photon energy, while clusters with a weaker non axial distortion give N1 line at higher photon energy, which are still at a lower photon energy than the R1 line of unperturbed Cr$^{3+}$ ions in axial symmetry.

From this EXAFS study, it is also possible to gain information about the nature of the dominant defects in the ZGO series. It has been previously shown that the less disordered ZGO material is d-ZGO, where the zinc deficiency in the starting composition compensates the Ga loss during high temperature synthesis.\cite{17} Increasing the Zn content in the synthesis increases the defect concentration (e-ZGO $>$ s-ZGO $>$ d-ZGO) as shown by the increasing intensity of N1 and N2 emission lines, presumably in the form of an excess of antisite defects ${Zn_{Ga}}'$.\cite{17} This hypothesis is now reinforced by the correlation between M-O bond length (M=Ga, Zn) and Cr-O bond length (Figure \ref{fig:corre} (c)). It is found that increasing the defect concentration results in an increase in Zn-O bond length and a decrease in Ga-O bond length (see Supplementary Table 4). Zn-O bond length is nearly equal to Ga-O bond length at the highest defect concentration (e-ZGO). This behaviour can be explained by the fact that increasing the number of Zn$^{2+}$ in Ga$^{3+}$ sites ($Zn_{Ga}$ antisite defects) has two effects: (i) as shown before, it lowers the symmetry of Cr$^{3+}$ site, which reduces the Cr-O bond length and shifts the $^2$E-$^4$A$_2$ emission to lower energy, and (ii) it introduces in the lattice more Zn-O bonds with the same length as Ga-O bonds, which shifts the average Zn-O bond length towards that of Ga-O as observed in Figure \ref{fig:corre} (c). The compensation for the excess of negative charge induced by ${Zn_{Ga}}'$ antisite defects should be insured by the other antisite defect ${Ga_{Zn}}^\circ$ and/or by oxygen vacancies ${V_{O}}^\circ$. In the case of compensation by ${Ga_{Zn}}^\circ$ antisite defect, we expect a shortening of the Ga-O bond upon increasing defect concentration. As there are two Ga for one Zn in the lattice, the ratio ${Ga_{Zn}}^\circ$/${Ga_{Ga}}^\times$ should be two times smaller than the ratio ${Zn_{Ga}}'$/${Zn_{Zn}}^\times$, where ${Ga_{Ga}}^\times$ and ${Zn_{Zn}}^\times$ represent Ga and Zn in their normal site. We thus expect a ratio R = -2 between the slopes of the Zn-O and Ga-O variations, which is close to the experimental value R $\approx$ -2.17 deduced from Figure \ref{fig:corre} (c). Thus the variations of Cr-O bond length with Zn-O and Ga-O bond lengths can be accounted for by $Zn_{Ga}$ and $Ga_{Zn}$ antisite defects present in the lattice. EPR spectroscopy previously indicated that, N2 emission line of LLP in ZGO might be due to a Cr$^{3+}$ ion with a ${Zn_{Ga}}'$ antisite defect in its first neighbour position (at 0.295 nm) and a ${Ga_{Zn}}^\circ$ antisite defect at slightly larger distances.\cite{17} This model is now supported by the present EXAFS study. It can also explain the origin of the considerable improvement of LLP in the gallogermanate series Zn$_{1+x}$Ga$_{2-2x}$Ge$_x$O$_4$.\cite{8} In these compounds, two Ga$^{3+}$ ions are replaced by one Ge$^{4+}$ ion and one Zn$^{2+}$ ion. Thus, increasing x should increase the number of Cr$^{3+}$ ions close to a negatively charged ${Zn_{Ga}}'$ and a positively charged ${Ge_{Ga}}^\circ$ defects. This defect configuration is determinant in our model for the visible light induced LLP mechanism.\cite{16,17}

\section{Conclusions}

The role of antisite defects on the persistent luminescence induced by visible light excitation in Cr$^{3+}$-doped AB$_2$O$_4$ spinels (A=Zn, Mg; B=Ga, Al) was investigated by a combined optical, EPR, XRD and EXAFS study. The main conclusions are the following. (i) Visible light excitation of persistent luminescence necessitates a small degree of inversion disorder, with the optimal level corresponding to that in d-ZGO. (ii) Increasing the defect concentration decreases the Cr-O bond length and the crystal field energy, attributed to an increasing $\pi$ bond contribution to the Cr-O interaction. (iii) Defects responsible for this Cr-O bond variation in ZGO are likely to be ${Zn_{Ga}}'$ and ${Ga_{Zn}}^\circ$ antisite defects. It can be concluded that a doping strategy which can control the amount of antisite defects should allow the optimization of the intensity and length of the persistent luminescence in AB$_2$O$_4$:Cr$^{3+}$ spinels.

\section{Acknowledgment}

Authors acknowledge SAMBA beamline, Soleil synchrotron facility, for giving the beamtime. Ms. St\'{e}phanie Belin, beamline scientist, is gratefully thanked for the experimental support. Financial support from Indo-French Centre for the Promotion of Advanced Research (IFCPAR)/ CEntre Franco-Indien Pour la Recherche Avanc\'{e}e (CEFIPRA) is acknowledged.


\end{document}